\documentclass[prd,showpacs,superscriptaddress,nofootinbib,floatfix,11pt]{revtex4-2}

\usepackage{geometry}
\usepackage{amsmath}
\usepackage[scr=boondox]{mathalpha}

\usepackage{array}

\usepackage{mathtools}
\usepackage{amssymb}
\usepackage{wasysym}
\usepackage{color}
\usepackage{slashed}
\usepackage{bbm}
\usepackage{graphicx}
\usepackage[caption=false]{subfig}
\usepackage{svg}
\usepackage[colorlinks=true,
breaklinks=true,
urlcolor=magenta,
citecolor=blue]{hyperref}
\usepackage{tikz}
\usepackage{multirow}
\usepackage{booktabs}
\usepackage{orcidlink}
\usepackage[normalem]{ulem}

\newcommand{\itp}{\affiliation{CAS Key Laboratory of Theoretical Physics, Institute of Theoretical Physics,\\ Chinese Academy of Sciences, Beijing 100190, China}}
\newcommand{\ucas}{\affiliation{School of Physical Sciences, University of Chinese Academy of Sciences, Beijing 100049, China}}
\newcommand{\peng}{\affiliation{Peng Huanwu Collaborative Center for Research and Education,\\ Beihang University, Beijing 100191, China}}
\newcommand{\scnt}{\affiliation{Southern Center for Nuclear-Science Theory (SCNT), Institute of Modern Physics,\\ Chinese Academy of Sciences, Huizhou 516000, China}}

\begin{document}

\title{Entanglement suppression and low-energy scattering of heavy mesons}

\author{Tao-Ran Hu\orcidlink{0009-0003-9720-0171}}\email{hutaoran21@mails.ucas.ac.cn}\ucas
\author{Su Chen}\email{chensu223@mails.ucas.ac.cn}\ucas
\author{Feng-Kun Guo\orcidlink{0000-0002-2919-2064}}\email{fkguo@itp.ac.cn}\itp\ucas\peng\scnt

\date{\today}

\begin{abstract}

Recently entanglement suppression was proposed to be one possible origin of emergent symmetries.
Here we test this conjecture in the context of heavy meson scatterings.
The low-energy interactions of $D^{(*)}\bar D^{(*)}$ and $D^{(*)} D^{(*)}$ are closely related to the hadronic molecular candidates $X(3872)$ and $T_{cc}(3875)^+$, respectively, and can be described by a nonrelativistic effective Lagrangian manifesting heavy-quark spin symmetry, which includes only constant contact potentials at leading order. We explore entanglement suppression in a tensor-product framework to treat both the isospin and spin degrees of freedom.  
Using the $X(3872)$ and $T_{cc}(3875)^+$ as inputs, we find that entanglement suppression indeed leads to an emergent symmetry, namely, a light-quark spin symmetry, and as such the $D^{(*)}\bar D^{(*)}$ or $D^{(*)} D^{(*)}$ interaction strengths for a given total isospin do not depend on the total angular momentum of light (anti)quarks.
The $X(3872)$ and $T_{cc}(3875)^+$ are predicted to have five and one isoscalar partner, respectively, while the corresponding partner numbers derived solely from heavy-quark spin symmetry are three and one, respectively.
The predictions need to be confronted with experimental data and lattice quantum chromodynamics results to further test the entanglement suppression conjecture.

\end{abstract}

\maketitle

\newpage

\section{Introduction}

Symmetries play a crucial role in physics, serving as fundamental principles for understanding Nature and revealing the properties of elementary particles and interactions. Particularly at low energies, the behavior of many systems can be attributed to the influence of their symmetries. Symmetries at low energies often manifest as local approximations of high-energy theories, and additional symmetries, known as ``emergent symmetries'' which are not in the action of the theory, may arise. In recent years, the concept of quantum entanglement has been introduced to the study and description of emergent symmetries, providing a new perspective for uncovering novel physical phenomena and understanding the behavior of low-energy systems~\cite{Beane:2018oxh, Low:2021ufv, Beane:2021zvo, Liu:2022grf, Carena:2023vjc, Liu:2023bnr, Kirchner:2023dvg}.

Entanglement measures the degree to which a system is entangled and quantifies the deviation from tensor-product structure for a given state~\cite{Einstein:1935rr, Bell:1964kc} (and if the tensor product structure is considered quasi-classical, then entanglement signifies the deviation from the classical structure). Given the ability to quantify the entanglement of a state, it is natural to extend this concept to quantify the entanglement of an operator~\cite{Zanardi:2001zza}. This can be achieved by averaging the entanglement measure of the states produced by applying the operator to all tensor-product states. It is evident that the entanglement of an operator measures its capacity to generate entanglement, termed as  ``entanglement power'' in the literature.

Since the $S$-matrix is also an operator, it is conceivable to assign an entanglement power to it. We know that the $S$-matrix carries all the information of a scattering process, hence its entanglement power measures the deviation of a specific scattering process from classical structure. It is natural to consider extreme cases, such as when the entanglement of the $S$-matrix reaches its maximum or minimum value (which can always be zero). 
Ref.~\cite{Beane:2018oxh} examines the latter, revealing some intriguing clues: the inherent $\text{SU}(2)_\text{isospin}\times\text{SU}(2)_\text{spin}$ symmetry of nucleon-nucleon scattering enlarges to Wigner's $\text{SU}(4)$ symmetry~\cite{Wigner:1936dx, Wigner:1937zz, Wigner:1939zz} when the entanglement vanishes, leading to the conjecture that entanglement suppression could be the origin of emergent symmetries. 

If the hypothesis holds, one may speculate that minimal entanglement would constrain the parameter space of low-energy hadron reactions, and thus would determine the emergence of new structures in hadronic interactions in the low-energy region. 
Consequences of entanglement suppression has been examined for nucleon-nucleon interactions~\cite{Beane:2018oxh,Low:2021ufv}, the pionic scattering~\cite{Beane:2021zvo}, the scattering between light octet baryons~\cite{Liu:2022grf,Liu:2023bnr}, and the relativistic scattering of Higgs doublets~\cite{Carena:2023vjc}.
In this paper, we will extend the study of entanglement suppression to the scattering of heavy mesons, where there are numerous intriguing near-threshold structures under intensive investigations. 
Two of the prominent examples in this regard are the $X(3872)$~\cite{Belle:2003nnu}, also known as the $\chi_{c1}(3872)$~\cite{ParticleDataGroup:2022pth}, and the $T_{cc}(3875)^+$~\cite{LHCb:2021vvq, LHCb:2021auc}, which have been proposed to be potential $D\bar D^*$~\cite{Close:2003sg,Pakvasa:2003ea,Voloshin:2003nt,Swanson:2003tb,Braaten:2003he,Tornqvist:2004qy} and $D^*D$~\cite{Li:2021zbw, Xin:2021wcr,Du:2021zzh,Albaladejo:2021vln} hadronic molecular states, respectively (for reviews, see Refs.~\cite{Hosaka:2016pey,Lebed:2016hpi,Esposito:2016noz,Guo:2017jvc,Olsen:2017bmm,Karliner:2017qhf,Kalashnikova:2018vkv,Brambilla:2019esw,Chen:2022asf,Meng:2022ozq}). 
We will investigate to what consequences their existence together with entanglement suppression can lead to.

The interaction between a pair of ground-state heavy mesons near threshold is closely related to formation of hadronic molecular states~\cite{Guo:2017jvc}. Therefore, we will analyze the near-threshold scattering processes of a pair of heavy mesons (namely, the $D^{(*)}D^{(*)}$ and $D^{(*)}\bar{D}^{(*)}$ scattering), where particles can be treated using a nonrelativistic approximation and the interaction is dominated by the lowest partial wave, i.e., the $S$-wave. Moreover, the mass of charm quark, $m_c$, is much larger than the nonperturbative energy scale of quantum chromodynamics (QCD), denoted by $\Lambda_\text{QCD}$. Therefore, when studying physical processes involving momentum scales of $\mathcal{O}(\Lambda_\text{QCD})$, we can treat $\Lambda_\text{QCD}/m_c$ as a small parameter and expand it in a power series to construct an effective field theory. The leading order (LO) is given by the heavy-quark limit ($m_c\rightarrow\infty$), where the heavy-quark spin symmetry (HQSS)~\cite{Isgur:1989vq} exists. HQSS has been used to predict heavy-quark spin partners of hadronic molecules containing heavy quark(s)~\cite{Guo:2009id,Voloshin:2011qa,Mehen:2011yh,Cleven:2015era}. This paper will investigate whether entanglement suppression will enlarge HQSS, then obtaining an emergent symmetry, which can predict more potential  siblings of $X(3872)$ and $T_{cc}(3875)^+$ than HQSS does.

Furthermore, enlarged symmetries in the low-energy region can also emerge in the large-$N_c$ limit, with $N_c$ the number of colors. For the Wigner's $\text{SU}(4)$ symmetry, the large-$N_c$ limit makes the same prediction, but for some other cases~\cite{Beane:2018oxh, Liu:2022grf}, the results obtained from the entanglement suppression and the large-$N_c$ limit differ, making it also meaningful to examine the differences between the two in the $D^{(*)}D^{(*)}$ and $D^{(*)}\bar{D}^{(*)}$ systems.

The outline of this paper is as follows: In Sec.~\ref{EP}, the entanglement power is considered in detail. The $S$-matrix is formulated in a basis convenient for calculation in Sec.~\ref{S}. Then in Sec.~\ref{SECERE}, we demonstrate how to relate the parameterization of the $S$-matrix to the amplitudes. In Sec.~\ref{EFT}, we present the effective Lagrangians and compute the amplitudes for $D^{(*)}D^{(*)}$ and $D^{(*)}\bar{D}^{(*)}$ scatterings. In Sec.~\ref{RE}, we derive the constraints imposed by entanglement suppression on the $S$-matrix, and in Sec.~\ref{DIS}, these results are connected to hadronic molecules, and we will show that with the $X(3872)$ and $T_{cc}(3875)^+$ as inputs, there emerges a light-quark spin symmetry. Subsequently, we conduct a brief large-$N_c$-limit analysis for the heavy meson scatterings in Sec.~\ref{LN}. Finally, a concise summary is provided in Sec.~\ref{SUM}.

\section{Entanglement power}\label{EP}

The degree to which a system is entangled, or its deviation from a tensor-product structure, provides a measure of how ``non-classical'' it is~\cite{Beane:2018oxh, Low:2021ufv}. An entanglement measure is a way to quantify the degree of entanglement of any given state. For a bipartite system $\left| \psi \right>$, the commonly employed linear entropy is defined as (see, e.g., Refs.~\cite{Beane:2021zvo, Liu:2022grf})
\begin{equation}
E(\left| \psi \right>) = 1 - \text{Tr}_1[\rho_1^2],
\end{equation}
where $\rho=\left| \psi \right> \left< \psi \right|$ is the density matrix, and $\rho_1 = \operatorname{Tr}_2(\rho)$ is the reduced density matrix obtained after tracing over subsystem 2. $E(\left| \psi \right>)$ serves as a
semi-positive definite measure of entanglement which vanishes only on tensor-product states $\left| \psi \right>=\left| \psi_1 \right>\otimes\left| \psi_2 \right>$, as shown in Appendix~\ref{app}.

Entanglement measure quantifies the entanglement in a quantum state $\left| \psi \right>$, while entanglement power measures the ability of a quantum-mechanical operator $U$ to generate entanglement by averaging over all states obtained by acting it on tensor-product states~\cite{Zanardi:2001zza}:
\begin{equation}
E(U) = \overline{E\left( U\left| \psi \right> \right) },\quad\left| \psi \right>=\left| \psi_1 \right>\otimes\left| \psi_2 \right>.
\end{equation}
By describing the average action of $U$ transiting a tensor-product state to an entangled state, entanglement power expresses a state-independent entanglement measure that is also semi-positive definite and vanishes, i.e., is minimized, only when $U\left| \psi \right>$ remains a tensor-product state for any $\left| \psi \right>=\left| \psi_1 \right>\otimes\left| \psi_2 \right>$.

In general, a low-energy scattering event can entangle position, spin, and other quantum numbers, and it is therefore natural to assign an entanglement power to the $S$-matrix for such a scattering process. Moreover, the small mass splitting between $u$ and $d$ quarks leads to the approximate $\text{SU}(2)$ isospin symmetry, so it is instructive to take into account the isospin invariance, which introduces interesting interplay between flavor and spin quantum numbers~\cite{Liu:2022grf}. Based on the above discussion, we choose to define the entanglement power of the $S$-matrix in the initial two-particle isospin $\otimes$ spin space. Since the $S$-wave heavy mesons are isospin-1/2 and spin-0 ($D$) or spin-1 ($D^*$), it is expedient to focus on the following two cases.

The first one is where there are two isospin states (a qubit) for each particle, the isospin-1/2 case.
This is just like the discussion of the spin states for the nucleon-nucleon case in Ref.~\cite{Beane:2021zvo}.
The most general initial isospin-1/2 state can be parameterized using the two complex parameters or four real parameters. 
Among them, one parameter can be removed by normalization and one gives an overall irrelevant phase. Finally only two real parameters are left, which parameterize a $\mathbb{CP}^1$ manifold, also known as the 2-sphere $S^2$ or the Bloch sphere~\cite{Beane:2021zvo,Bengtsson:2006rfv}. It can be parameterized as 
\begin{align}
	\left| \psi \right>= \left(\cos\frac{\theta}{2},e^{i\phi}\sin\frac{\theta}{2}\right)^T,
\end{align}
with $\theta\in[0,\pi]$ and $\phi\in[0,2\pi)$. 
Therefore, the incoming state of two isospin-1/2 particles is mapped to a point on the product manifold, $\mathbb{CP}^1\times\mathbb{CP}^1$ , while the entanglement power $E(S)$ of the $S$-matrix is defined as
\begin{equation}\label{Eq.SU(2)}
E(S)=1-\int\frac{\mathrm{d}\Omega_1}{4\pi}\frac{\mathrm{d}\Omega_2}{4\pi}\mathrm{Tr}_1[\rho_1^2],
\end{equation}
where we have defined $
\rho =|\psi _{\mathrm{out}}\rangle \langle \psi _{\mathrm{out}}|$ and $|\psi _{\mathrm{out}}\rangle =S|\psi _{\mathrm{in}}\rangle$, ~$|\psi _{\mathrm{in}}\rangle =|\psi _1\rangle \otimes |\psi _2\rangle $.

In the spin-1 case, we have three spin states (a qutrit) which involve three complex parameters, similar to the isospin space of the $\pi\pi$ scattering discussed in Ref.~\cite{Beane:2021zvo}. 
Four real parameters are left after considering normalization and removing the overall phase, and they parameterize the $\mathbb{CP}^2$ manifold. Thus, an arbitrary qutrit can be written as
\begin{equation}
\left| \psi \right>=(\cos\beta\sin\alpha,e^{i\mu}\sin\beta\sin\alpha,e^{i\nu}\cos\alpha)^T,
\end{equation}
where $\alpha,\beta\in[0,\pi/2]$ and $\mu,\nu\in[0,2\pi)$. Similarly to Eq.~\eqref{Eq.SU(2)}, the entanglement power $E(S)$ of the $S$-matrix can be defined as
\begin{equation}\label{Eq.SO(3)}
E(S)=1-\int\mathrm{d}\omega_1\mathrm{d}\omega_2\mathrm{Tr}_1[\rho_1^2],
\end{equation}
with $\mathrm{d}\omega=(2/\pi^2)\cos\alpha\sin^3\alpha\mathrm{d}\alpha\cos\beta\sin\beta\mathrm{d}\beta\mathrm{d}\mu\mathrm{d}\nu$ the normalized measure that describes the geometry of $\mathbb{CP}^2$~\cite{Bengtsson:2001yd, Bengtsson:2006rfv}.

\section{Heavy meson scattering}

\subsection{$S$-matrix}\label{S}

In this paper, we primarily study the scattering of heavy mesons in the near-threshold region, which is dominated by the $S$-wave interaction. In general, the $S$-matrix can be expressed as
\begin{equation}
S=\sum_{I,J}{\mathcal{J}_{J}\otimes \mathcal{I}_{I}}\, e^{2i\delta _{IJ}},
\end{equation}
where we define $\mathcal{J}_{J}\otimes \mathcal{I}_{I}$ the projection operators onto subspaces of definite isospin $I$ and total spin $J$, and $\delta _{IJ}$ the corresponding phase shift.

Let us start with $D^{(*)}D^{(*)}$ scattering. The construction of the $S$-matrix proceeds straightforwardly~\cite{Beane:2021zvo, Liu:2022grf}:
\begin{align}
	S_{DD} &= \mathcal{I}_0\, e^{2i\delta_{00}} + \mathcal{I}_1 \, e^{2i\delta_{10}}, \\
	S_{D^*D} &= \mathcal{I}_0\, e^{2i\delta_{01}} + \mathcal{I}_1\, e^{2i\delta_{11}}, \\
	S_{D^*D^*} &= \sum_{I=0,1} \sum_{J=0,1,2} \mathcal{I}_I \otimes \mathcal{J}_J\, e^{2i\delta_{IJ*}}, \label{eq:SD*D*}
\end{align}
with the isospin space projectors
\begin{equation}
	\mathcal{I}_0\equiv \frac{1-\boldsymbol{\tau }_1\cdot \boldsymbol{\tau }_2}{4}, \quad 	
	\mathcal{I}_1 \equiv \frac{3+\boldsymbol{\tau }_1\cdot \boldsymbol{\tau }_2}{4}, 
	\label{eq:Iprojectors}
\end{equation}
and the spin space projectors
\begin{equation}
\begin{aligned}
	\mathcal{J}_0 &\equiv -\frac{1}{3}\left[ 1-\left( \boldsymbol{t}_1\cdot \boldsymbol{t}_2 \right) ^2 \right],\\
	\mathcal{J}_1 &\equiv 1-\frac{1}{2}(\boldsymbol{t}_1\cdot \boldsymbol{t}_2)-\frac{1}{2}\left( \boldsymbol{t}_1\cdot \boldsymbol{t}_2 \right) ^2 , \\
	\mathcal{J}_2 &\equiv \frac{1}{3}\left[ 1+\frac{3}{2}\left( \boldsymbol{t}_1\cdot \boldsymbol{t}_2 \right) +\frac{1}{2}\left( \boldsymbol{t}_1\cdot \boldsymbol{t}_2 \right) ^2 \right],
\end{aligned}
\end{equation}
where $(t_{1,2}^a)^{bc}=-i\epsilon^{abc}$, $\boldsymbol{\tau}$ are Pauli matrices in the flavor space, and the $D^*D^*$ scattering phase shifts are denoted as $\delta _{IJ*}$,
to be distinguished from the $D^{(*)} D$ scattering phase shifts $\delta _{IJ}$.
The $S$-matrices for $DD$ and $D^{*}D$ scattering are exclusively parameterized in the isospin space. This is because in these two processes, the total spin has only one specific value for each. Additionally, the Bose-Einstein statistics dictates that: (i) $\delta_{00}=0$ for $DD$ scattering; (ii) $\delta_{00*} = \delta_{02*} =  \delta_{11*} =0$ for $D^{*}D^{*}$ scattering, i.e., the total isospin $I = 0$ projects into spin-triplet $^3S_1$ while $I = 1$ projects into both spin-singlet $^1S_0$ and quintuplet $^5S_2$~\cite{Du:2021zzh}.

For $D^{(*)}\bar{D}^{(*)}$ scattering, there are two additional intricacies: (i) electrically neutral $D^{(*)}\bar{D}^{(*)}$ combinations should have definite $C$-parities; (ii) there is no Bose-Einstein statistics. This implies that the $S$-matrix for $D^{(*)}\bar{D}^{(*)}$ scattering should be written as
\begin{align}
S_{D\bar{D}}&=\mathcal{I}_0e^{2i\bar{\delta}_{00}}+\mathcal{I}_1e^{2i\bar{\delta}_{10}},\\
\label{pm}
S_{D\bar{D}^*\pm}&=\mathcal{I}_0e^{2i\bar{\delta}_{01\pm}}+\mathcal{I}_1e^{2i\bar{\delta}_{11\pm}},\\
S_{D^*\bar{D}^*}&=\sum_{I=0,1} \sum_{J=0,1,2} \mathcal{I}_I \otimes \mathcal{J}_J\, e^{2i\bar{\delta}_{IJ*}}.
\end{align}
The phase shifts of $D\bar{D}$ and $D^*\bar D^*$ scatterings are denoted as $\bar{\delta} _{IJ}$ and $\bar{\delta} _{IJ*}$, respectively. The $J^{PC}$ combinations that a pair of $D^{(*)}$ and $\bar{D}^{(*)}$ can form are as follows~\cite{Nieves:2012tt,Guo:2017jvc}:
\begin{equation}\label{eq:HHcoms}
\begin{aligned}
	0^{++}:&\quad D\bar{D},~ D^*\bar{D}^*;\\
	1^{+-}:&\quad \frac{1}{\sqrt{2}}\left( D\bar{D}^*+D^*\bar{D} \right) ,~ D^*\bar{D}^*;\\
	1^{++}:&\quad \frac{1}{\sqrt{2}}\left( D\bar{D}^*-D^*\bar{D} \right) ;\\
	2^{++}:&\quad D^*\bar{D}^*.
\end{aligned} 
\end{equation}
Thus, it is seen in Eq.~\eqref{pm} that there are two independent $S$-matrices in $D\bar{D}^*$ channel, with $C=\pm$ and thus the corresponding subindex ``$_\pm$''.
Here the phase convention for the charge conjugation is chosen as $\hat C\left| D\right>=\left| \bar{D}\right>$ and $\hat C\left| D^*\right>=-\left| \bar{D}^*\right>$.

\subsection{Effective range expansion}\label{SECERE}

In this subsection, we will first derive the effective range expansion that will be utilized later, and then discuss how to relate phase shifts to amplitudes in different physical cases.

We start by considering the effective Lagrangian for two nonrelativistic spinless bosons $\phi_i$ with only the LO contact interaction in a derivative (nonrelativistic) expansion:
\begin{equation}
\mathcal{L}=\sum_{i=1,2}\phi_i^\dagger\left(i\partial_t-m_i+\frac{\nabla^2}{2m_i}\right)\phi_i-{C_0}\phi_1^\dagger\phi_2^\dagger\phi_1\phi_2.
\end{equation}

\begin{figure}
    \centering
    \includegraphics[width=0.75\textwidth]{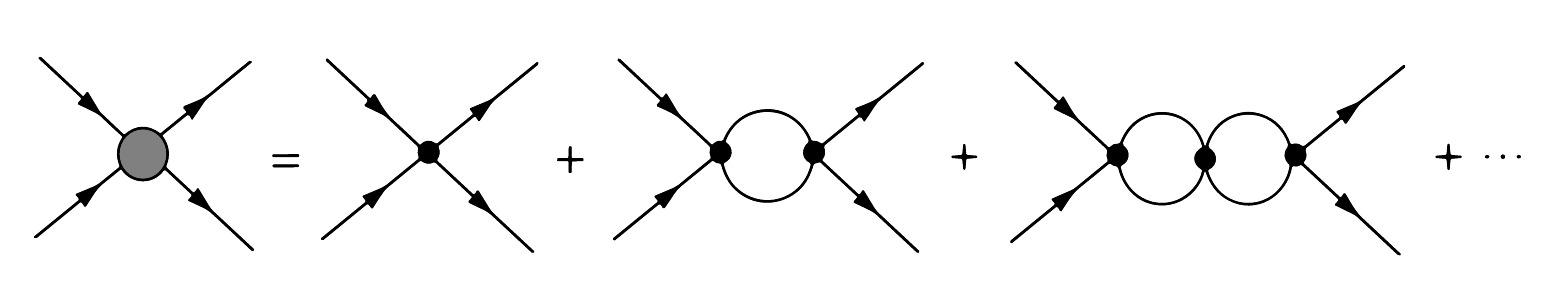}
    \caption{The first few diagrams contributing to the $S$-wave amplitude for the process $\phi_1\phi_2\rightarrow\phi_1\phi_2$. The solid black dot represents the $-iC_0$ vertex.}
    \label{fig: Diagrams}
\end{figure}

It is easy to write down the amplitude for the process $\phi_1\phi_2\rightarrow\phi_1\phi_2$ shown in Fig.~\ref{fig: Diagrams}:
\begin{equation}\label{Amp}
i\mathcal{M} =-iC_0+\left( -iC_0 \right) \left( iG \right) \left( -iC_0 \right) +\cdots=-\frac{i}{C_{0}^{-1}-G},
\end{equation}
where we define the two-point loop function
\begin{equation}
G=i\int{\frac{\mathrm{d}k^0\mathrm{d}^3\boldsymbol{k}}{(2\pi )^4}}\left[ \left( k^0-\frac{\boldsymbol{k}^2}{2m_1}+i\epsilon \right) \left( E-k^0-\frac{\boldsymbol{k}^2}{2m_2}+i\epsilon \right) \right] ^{-1}=-\frac{\mu}{2\pi}\left( \Lambda +ip \right),
\end{equation}
with $m_1$ and $m_2$ the boson masses, $\mu$ the reduced mass, $p=\sqrt{2\mu E}$ the magnitude of the center-of-mass momentum, and $\Lambda$ the cut-off introduced to regularize the loop integral. 

Meanwhile, the $S$-matrix for nonrelativistic elastic scattering can be written as
\begin{equation}
S=e^{2i\delta}=1+\frac{\mu p}{\pi}i\mathcal{M} ,
\end{equation}
where $\delta$ is the phase shift. Hence, one has
\begin{equation}\label{Mex}
i\mathcal{M} =\frac{2\pi}{\mu}\frac{i}{p\cot \delta -ip},
\end{equation}
which, combined with Eq.~\eqref{Amp}, directly yields 
\begin{equation}\label{ERE}
p\cot \delta =-\frac{2\pi}{\mu C_0}-\Lambda,
\end{equation}
which is the effective range expansion~\cite{Bethe:1949yr} at LO (see, e.g., Refs.~\cite{Kaplan:1996xu,Hammer:2019poc}), with $a = \left(2\pi/(\mu C_0)+\Lambda\right)^{-1}$  the $S$-wave scattering length.

Based on the above discussion, we can deduce the contact potential $C_0$ for certain phase shift values. In particular, the noninteracting and unitary limit cases are of utmost importance at LO:
\begin{align}
\delta=0:&\quad a = 0~~\,\text{or}~ C_0=0,\\
\delta=\frac\pi 2:&\quad a=\infty~\text{or}~C_0=-\frac{2\pi}{\mu\Lambda}.
\end{align}
In fact, at LO only these two limits are momentum-independent since the phase shift $\delta(p)$ is in general a function of $p$, so if the entanglement suppression constraint (see Sec.~\ref{RE}) is enforced at one value of momentum it will generically not hold at other values of momentum. 
Additionally, both the free theory ($\delta=0$) and the theory at the unitarity limit ($\delta=\pi/2$) are invariant under the Schrödinger symmetry~\cite{Mehen:1999nd,Low:2021ufv}, which is the nonrelativistic conformal group and the largest symmetry group preserving the Schrödinger equation. They correspond to the two fixed points of renormalization group running of nonrelativistic two-body scattering by a short-range potential~\cite{Birse:1998dk}.

\subsection{LO effective field theory for heavy meson scattering}\label{EFT}

At very low energies, the LO $D^{(*)}D^{(*)}$ interaction in the nonrelativistic effective field theory follows from the effective Lagrangian which contains only constant contact potentials~\cite{Fleming:2021wmk,Du:2021zzh},
\begin{equation}\label{Eq.HH}
\begin{aligned}
\mathcal{L}_{HH}=&-\frac{D_{00}}8\mathrm{Tr}\left[H^{a\dagger}H_{b}H^{b\dagger}H_{a}\right]-\frac{D_{01}}8\mathrm{Tr}\left[H^{a\dagger}H_{b}\sigma^{m}H^{b\dagger}H_{a}\sigma^{m}\right]\\
&-\frac{D_{10}}8\mathrm{Tr}\left[H^{a\dagger}H_bH^{c\dagger}H_d\right]\boldsymbol{\tau}_{a}^{d}\cdot\boldsymbol{\tau}_{c}^{b}-\frac{D_{11}}8\mathrm{Tr}\left[H^{a\dagger}H_{b}\sigma^{m}H^{c\dagger}H_{d}\sigma^{m}\right]\boldsymbol{\tau}_{a}^{d}\cdot\boldsymbol{\tau}_{c}^{b},
\end{aligned}
\end{equation}
where $\boldsymbol{\sigma}$ denotes the Pauli matrices in the $\text{SU}(2)$ spinor space, $D_{00,01,10,11}$ are light-flavor-independent low-energy constants (LECs), $\operatorname{Tr}[\cdot]$ takes trace in the spinor space, and $\boldsymbol{\tau}\cdot\boldsymbol{\tau}$ sums over all Pauli matrices in the flavor space. The above Lagrangian respects both HQSS and isospin symmetry.

Moreover, the LO Lagrangian for the low-energy $S$-wave interaction between a pair of heavy and anti-heavy mesons containing only constant contact terms reads~\cite{AlFiky:2005jd,Nieves:2012tt,Ji:2022uie}\footnote{One can check that the double-trace form can also be rewritten in the single-trace form like Eq.~\eqref{Eq.HH} using the completeness relation for the Pauli matrices, $2\delta_i^l \delta_k^j= \delta_i^j \delta_k^l+ \boldsymbol{\sigma}_i^j \cdot \boldsymbol{\sigma}_k^l$, as shown in~\cite{Mehen:2011yh}.}
\begin{equation}\label{HHbarL}
\begin{aligned}
\mathcal{L}_{H\bar{H}}=&-\frac{1}{4}\operatorname{Tr}\left[H^{a\dagger}H_{b}\right]\operatorname{Tr}\left[\bar{H}^{c}\bar{H}_{d}^{\dagger}\right]\left(F_{A}\delta_{a}^{b}\delta_{c}^{d}+F_{A}^{\tau}\boldsymbol{\tau}_{a}^{b}\cdot\boldsymbol{\tau}_{c}^{d}\right)\\
&+\frac{1}{4}\operatorname{Tr}\left[H^{a\dagger}H_{b}\sigma^{m}\right]\operatorname{Tr}\left[\bar{H}^{c}\bar{H}_{d}^{\dagger}\sigma^{m}\right]\left(F_{B}\delta_{a}^{b}\delta_{c}^{d}+F_{B}^{\tau}\boldsymbol{\tau}_{a}^{b}\cdot\boldsymbol{\tau}_{c}^{d}\right).
\end{aligned}
\end{equation}
Again, there are four light-flavor-independent LECs $F_{A,B}^{(\tau)}$.

In the above Lagrangians, the heavy and anti-heavy mesons are grouped into superfields as~\cite{Manohar:2000dt,Hu:2005gf}
\begin{equation}
H_a=P_a+\boldsymbol{P}^*_a\cdot\boldsymbol{\sigma},\quad\bar{H}^a=\bar{P}^a+\bar{\boldsymbol{P}}^{*a}\cdot\boldsymbol{\sigma},
\end{equation}
with $P_a$ and $\boldsymbol{P}^*_a$ annihilating the ground-state pseudoscalar and vector charmed mesons, respectively, and $\bar{P}^a$ and $\bar{\boldsymbol{P}}^{*a}$ annihilating the anti-charmed mesons.

The contact potentials for the different isospin/spin-parity $D^{(*)}D^{(*)}$ $S$-wave channels derived from the Lagrangian of Eq.~\eqref{Eq.HH} read~\cite{Du:2021zzh}
\begin{align}
T^{IJ=00}\left( DD \right) &=\frac{1}{2}(D_{00}+3D_{01}+D_{10}+3D_{11}),\\
\label{Tcc+}
T^{IJ=01}\left( D^*D \right) &=-2(D_{01}-3D_{11}),\\
T^{IJ=11}\left( D^*D \right) &=D_{00}+D_{01}+D_{10}+D_{11},\\
\label{Tcc+*}
T^{IJ=01}\left( D^*D^* \right) &=-2(D_{01}-3D_{11}),\\
T^{IJ=10}\left( D^*D^* \right) &=-\frac{1}{2}(D_{00}-5D_{01}+D_{10}-5D_{11}),\\
T^{IJ=12}\left( D^*D^* \right) &=D_{00}+D_{01}+D_{10}+D_{11}.
\end{align}
All other potentials vanish, where we have used the fact that the isoscalar (isovector) wave function for two identical particles with isospin $I=1/2$ are antisymmetric (symmetric), respectively.

When it comes to the $D^{(*)}\bar{D}^{(*)}$ case, the four LECs that appear in Eq.~\eqref{HHbarL} are often rewritten for convenience into $C_{0a}$, $C_{0b}$ and $C_{1a}$, $C_{1b}$~\cite{Hidalgo-Duque:2012rqv}, which stand for the LECs in the isospin $I = 0$ and $I=1$ channels, respectively. The relations read
\begin{equation}
\begin{aligned}
C_{0a}&=F_A+\frac{10}3F_A^\tau,\quad C_{1a}=F_A-\frac23F_A^\tau,\\C_{0b}&=F_B+\frac{10}3F_B^\tau,\quad C_{1b}=F_B-\frac23F_B^\tau.
\end{aligned} \label{eq:LECs_C}
\end{equation}
Then the contact potentials can be expressed utilizing these four LECs~\cite{Ji:2022uie}:
\begin{align}
T^{IJ=00}\left( D\bar{D} \right) =C_{0a},&\quad T^{IJ=10}\left( D\bar{D} \right) =C_{1a},\\
T_{-}^{IJ=01}\left( D\bar{D}^* \right)=C_{0a}-C_{0b},&\quad T_{-}^{IJ=11}\left( D\bar{D}^* \right)=C_{1a}-C_{1b},\\
\label{X}
T_{+}^{IJ=01}\left( D\bar{D}^* \right)=C_{0a}+C_{0b},&\quad T_{+}^{IJ=11}\left( D\bar{D}^* \right)=C_{1a}+C_{1b},\\
T^{IJ=00}\left( D^*\bar{D}^* \right) =C_{0a}-2C_{0b},&\quad T^{IJ=10}\left( D^*\bar{D}^* \right) =C_{1a}-2C_{1b},\\
 T^{IJ=01}\left( D^*\bar{D}^* \right) =C_{0a}-C_{0b},&\quad T^{IJ=11}\left( D^*\bar{D}^* \right) =C_{1a}-C_{1b},\\
\label{Xp}
T^{IJ=02}\left( D^*\bar{D}^* \right) =C_{0a}+C_{0b},&\quad T^{IJ=12}\left( D^*\bar{D}^* \right) =C_{1a}+C_{1b},
\end{align}
where the lower indices ``$_\pm$'' represent $(D\bar D^*\mp D^*\bar D)/\sqrt{2}$ with different $C$-parities in Eq.~\eqref{eq:HHcoms}.

\section{Results}\label{RE}

Having set up the theoretical framework, we can now calculate the entanglement power and require it to vanish, which gives constraints on phase shifts. In this way, we can check the consequences of the constraints on amplitudes, which lead to relations among the LECs in the Lagrangian. It is not difficult for scatterings involving pseudoscalar mesons, as entanglement occurs solely in the isospin space, while for $D^*D^*$ and $D^*\bar{D}^*$ scatterings, it is useful to express what minimal entanglement means in a tensor-product space. Clearly, entanglement being zero in a large space is equivalent to it being zero in all of its subspaces. For instance, applying $\mathcal{I}_{I^\prime}\equiv \mathcal{I}_{I^\prime}\otimes1$ to $S=\sum_{I,J}\mathcal{I}_I \otimes \mathcal{J}_J\, e^{2i\delta_{IJ*}}$ gives
\begin{equation}
\mathcal{I}_{I^\prime} S=\sum_{I,J}{\mathcal{I}_{I^\prime}\mathcal{I}_{I}\otimes \mathcal{J}_{J}}e^{2i\delta _{IJ*}}=\mathcal{I}_{I^\prime}\otimes\sum_{J}{ \mathcal{J}_{J}}e^{2i\delta _{IJ^\prime*}} \equiv \mathcal{I}_{I^\prime} \otimes S_{I^\prime}.
\end{equation}
So the vanishing of entanglement implies that the entanglement vanishes in this spin subspace with isospin $I^\prime$. Similarly, for any isospin subspace with a specific total spin, the entanglement should also be zero.
Nucleon-nucleon scattering is also a process entangled in both spin and isospin spaces. In Ref.~\cite{Beane:2018oxh}, the parameterization was carried out only in spin space. This is because Fermi-Dirac statistics results in entanglement effectively occurring in only one space. Similarly, we find that in $D^{(*)}D^{(*)}$ scattering, spin entanglement and isospin entanglement also yield completely consistent results, so it is actually sufficient to compute entanglement in only one space. However, for $D^{(*)}\bar{D}^{(*)}$ scattering, there is no Bose-Einstein statistics, so this tensor-product structure is necessary, and such a formalism can be extended to cases entangling more quantum numbers.

Based on the above discussion, we only need to compute the entanglement power in two scenarios using Eqs.~\eqref{Eq.SU(2)} and \eqref{Eq.SO(3)}.
At LO of the heavy quark expansion, the $D$ and $D^*$ masses are the same. We also consider the isospin symmetric limit such that charged and neutral mesons in the same isospin multiplet are degenerate. Thus we will take $\mu=M/2$ with $M$ denoting the charmed meson mass in the following.

We start with the $S$-matrix in the isospin subspace with a specific total spin $J$ (both particles are isospin-1/2 states):
\begin{equation}
S_J=\mathcal{I}_0e^{2i\delta _{0J}}+\mathcal{I}_1e^{2i\delta _{1J}}.
\end{equation}
The Bose-Einstein forbidden cases of $DD$ and $D^*D^*$ scatterings are formally included by requiring $\delta _{00}=0$ and $\delta _{00*}=\delta_{02*}=\delta_{11*}=0$ (recall that in Eq.~\eqref{eq:SD*D*} we have introduced the ``$_*$'' subindex for vector-vector scattering phase shifts), respectively.
Evaluating the entanglement power using Eq.~\eqref{Eq.SU(2)} yields
\begin{equation}
E(S_J)=\frac16\sin^2[2(\delta_{0J}-\delta_{1J})],
\end{equation}
which vanishes only when
\begin{align}
	\left|\delta_{0J}-\delta_{1J}\right|=0~~\text{or}~~\frac\pi 2.
\end{align}
The solutions to equation $\left|\delta_{0J}-\delta_{1J}\right|=\pi/2$, namely $\delta_{0J} = 0$ and $\delta_{1J}=\pi/2$ (or vice versa), correspond to the cases of no interaction and the unitarity limit, respectively.
The latter is equivalent to taking the limit of exactly infinite
scattering length.

For vector meson scatterings, one also needs to consider the $S$-matrix in the spin subspace with a specific isospin $I$:
\begin{equation}
S_I=\mathcal{J}_0e^{2i\delta_{I0*}}+\mathcal{J}_1e^{2i\delta_{I1*}}+\mathcal{J}_2 e^{2i\delta_{I2*}}.
\end{equation}
Again, Bose-Einstein statistics requires $\delta_{00 *}=\delta_{02 *}=\delta_{11 *}=0$ for $D^* D^*$ scattering, while it does not constrain anything for $D^*\bar D^*$ scattering, as mentioned above.
The entanglement power can be calculated using Eq.~\eqref{Eq.SO(3)} and reads
\begin{equation}
\begin{aligned}
E(S_I) = &\,\frac{1}{648}\Big\{ 156 - 6\cos[4(\delta_{I0*}-\delta_{I1*})] - 65\cos[2(\delta_{I0*}-\delta_{I2*})] \\
&- 10\cos[4(\delta_{I0*}-\delta_{I2*})] - 60\cos[4(\delta_{I2*}-\delta_{I1*})] - 15\cos[2(\delta_{I0*}+\delta_{I2*}-2\delta_{I1*})] \Big\},
\end{aligned}
\end{equation}
which has only two non-entangling solutions:
\begin{equation}
\left|\delta_{I0*}-\delta_{I1*}\right|=\left|\delta_{I2*}-\delta_{I1*}\right|=0 ~~\text{or}~~\frac\pi 2.
\end{equation}

The subsequent step involves relating these solutions to the amplitudes using Eq.~\eqref{ERE}, which can be then be confronted to experimental or lattice QCD results, or use such empirical results as further input to select solutions and explore their implications.

\section{Consequences on heavy-meson hadronic molecules}\label{DIS}

It is already known that the near-threshold interaction between a pair of ground-state heavy mesons is closely related to formation of hadronic molecular states~\cite{Guo:2017jvc}. The $X(3872)$~\cite{Belle:2003nnu} has been proposed as a candidate of an isoscalar $D\bar{D}^*$ hadronic molecule with $J^{PC}=1^{++}$ quantum numbers~\cite{LHCb:2013kgk} for a long while~~\cite{Close:2003sg,Pakvasa:2003ea,Voloshin:2003nt,Swanson:2003tb,Braaten:2003he,Tornqvist:2004qy}. Moreover, in 2021 the LHCb Collaboration announced the discovery of $T_{cc}(3875)^+$ with preferred quantum numbers $I(J^P)=0(1^+)$~\cite{LHCb:2021vvq,LHCb:2021auc}, a double-charm $D^*D$ molecular candidate~\cite{Li:2021zbw, Xin:2021wcr,Du:2021zzh,Albaladejo:2021vln}, which reveals itself as a high-significance peaking structure in the $D^0D^0\pi^+$ and $D^+D^0\pi^0$ invariant mass distributions just below the nominal $D^{*+}D^0$ threshold. 
The masses of these two particles are extremely close to the $D^{0}\bar D^{*0}$ and $D^{*+}D^0$ thresholds, respectively,
\begin{align}
	M_{X} - M_{D^0} - M_{\bar D^{*0}} = 0.00_{-0.15}^{+0.09}\,\text{MeV}, ~
	M_{T_{cc}^+} - M_{D^{*+}} - M_{D^0} = (-0.36\pm0.04)\,\text{MeV},
\end{align}
where we have used the charmed meson masses from Ref.~\cite{ParticleDataGroup:2022pth}, the $X(3872)$ mass from the Flatt\'e analysis in Ref.~\cite{LHCb:2020xds}, and the $T_{cc}(3875)^+$ mass from the coupled-channel analysis with full $DD\pi$ three-body effects in Ref.~\cite{Du:2021zzh}. 

The existence of the isoscalar $X(3872)$ and $T_{cc}(3875)^+$ states so close to the $D\bar D^*$ and $D^* D$ thresholds, respectively, implies that the near-threshold $S$-wave interactions in both channels approach the unitary limit, with the corresponding $S$-wave scattering lengths being infinitely large. By taking these conditions as input, the entanglement suppression solutions can be further pinned down.
Consequently, partners of the $X(3872)$ and $T_{cc}(3875)^+$ states can be predicted. 
If some of these partners are not predicted by the intrinsic HQSS, one can assert that they arise from an emergent symmetry dictated by entanglement suppression. 

For $D^{(*)}D^{(*)}$ scattering, the $T_{cc}(3875)^+$ implies $\delta_{01}=\pi/2$. Then one obtains two solutions:
\begin{equation}\label{Eq.HHphaseshift1}
\delta _{01}=\delta _{01*}=\frac{\pi}{2},\quad\delta _{10}=\delta _{11}=\delta _{10*}=\delta _{12*}=0,
\end{equation}
or
\begin{equation}\label{Eq.HHphaseshift2}
\delta _{01}=\delta _{01*}=\delta _{10}=\delta _{11}=\delta _{10*}=\delta _{12*}=\frac{\pi}{2}.
\end{equation}

\begin{table}[tb]
\centering
\caption{Partners of the $T_{cc}(3875)^+$ predicted by HQSS or the two solutions of entanglement suppression given in Eqs.~\eqref{Eq.HHphaseshift1} and \eqref{Eq.HHphaseshift2}. The symbol ``$\odot$'' denotes the input $T_{cc}(3875)^+$ state, ``$\otimes$'' represents its predicted partners, ``$\oslash$'' indicates that no near-threshold state is allowed, ``$\ocircle$'' is forbidden by Bose-Einstein statistics, and ``$-$'' signifies that no prediction can be made without further inputs.}
\begin{ruledtabular}
\begin{tabular}{p{5em} *{6}{c}}
\multirow{2}{*}{Channel} & \multicolumn{2}{c}{HQSS} & \multicolumn{2}{c}{Eq.~\eqref{Eq.HHphaseshift1} predictions} & \multicolumn{2}{c}{Eq.~\eqref{Eq.HHphaseshift2}  predictions} \\
\cmidrule(lr){2-3}\cmidrule(lr){4-5}\cmidrule(lr){6-7}
& $\quad I=0 \quad$ & $\quad I=1 \quad$ & $\quad I=0 \quad$ & $\quad I=1 \quad$ & $\quad I=0 \quad$ & $\quad I=1 \quad$ \\
\midrule
$DD(0^{+})$ & $\ocircle$ & $-$ & $\ocircle$ & $\oslash$ & $\ocircle$ & $\otimes$ \\
$D^*D(1^{+})$ & $\odot$ & $-$ & $\odot$ & $\oslash$ & $\odot$ & $\otimes$ \\
$D^*D^*(0^{+})$ & $\ocircle$ & $-$ & $\ocircle$ & $\oslash$ & $\ocircle$ & $\otimes$ \\
$D^*D^*(1^{+})$ & $\otimes$ & $\ocircle$ & $\otimes$ & $\ocircle$ & $\otimes$ & $\ocircle$ \\
$D^*D^*(2^{+})$ & $\ocircle$ & $-$ & $\ocircle$ & $\oslash$ & $\ocircle$ & $\otimes$
\end{tabular}
\end{ruledtabular}
\label{HHtab}
\end{table}

In both scenarios, an additional $D^*D^*$ zero-energy bound molecular state in the isoscalar $J^P=1^+$ sector, $T_{cc}^{*+}$, can be predicted based on Eqs.~\eqref{Eq.HHphaseshift1} and \eqref{Eq.HHphaseshift2}, i.e., $\delta_{01*}=\pi/2$. However, it is not a result of entanglement suppression but stems from HQSS~\cite{Albaladejo:2021vln, Du:2021zzh}, as can be seen from $T^{IJ=01}\left( D^*D \right)=T^{IJ=01}\left( D^*D^* \right)$ in Eqs.~\eqref{Tcc+} and \eqref{Tcc+*}. 
The additional consequences of entanglement suppression is that the interaction strengths of the isovector channels are all the same, either noninteracting as in Eq.~\eqref{Eq.HHphaseshift1} or at the unitary limit as in Eq.~\eqref{Eq.HHphaseshift2}. In the latter instance, we would also anticipate four extra weakly bound states near the $D^{(*)}D^{(*)}$ threshold. The results are shown in Table~\ref{HHtab}.
In both cases, it means that the symmetry for the spin degree of freedom of the light quarks in the heavy meson pair is enlarged from SU(2)$\times$SU(2) to SU(4).

It is also instructive to explicitly write out the solution of the LECs for the first case~\eqref{Eq.HHphaseshift1}:
\begin{equation}
D_{00}+D_{10}=0,\quad D_{01}=\frac{\pi}{4\mu \Lambda},\quad D_{11}=-\frac{\pi}{4\mu \Lambda},
\end{equation}
which yields the Lagrangian as
\begin{equation}\label{lag}
\begin{aligned}
	\mathcal{L} _{HH}=&-\frac{D_{00}}{8}\mathrm{Tr}\left[ H^{a\dagger}H_bH^{b\dagger}H_a \right] -\frac{\pi}{32\mu \Lambda}\mathrm{Tr}\left[ H^{a\dagger}H_b\sigma ^mH^{b\dagger}H_a\sigma ^m \right]\\
	&+\frac{D_{00}}{8}\mathrm{Tr}\left[ H^{a\dagger}H_bH^{c\dagger}H_d \right] \boldsymbol{\tau }_{a}^{d}\cdot \boldsymbol{\tau }_{c}^{b}+\frac{\pi}{32\mu \Lambda}\mathrm{Tr}\left[ H^{a\dagger}H_b\sigma ^mH^{c\dagger}H_d\sigma ^m \right] \boldsymbol{\tau }_{a}^{d}\cdot \boldsymbol{\tau }_{c}^{b}.\\
\end{aligned}
\end{equation}
It is seen in Eq.~\eqref{Eq.HHphaseshift1} that all amplitudes for isovector channels vanish, in agreement with the Lagrangian~\eqref{lag} proportional to the projector onto isoscalar subspace, $\mathcal{I}_0=(1-\boldsymbol{\tau }_1\cdot \boldsymbol{\tau }_2)/4$.

For $D^{(*)}\bar{D}^{(*)}$ scattering, the $X(3872)$ implies $\bar{\delta}_{01+}=\pi/2$, where we use $\bar\delta$ to denote $D^{(*)}\bar D^{(*)}$ scattering phase shifts and the subscript ``$_+$'' denotes the positive $C$-parity combination of $D\bar D^*$ in Eq.~\eqref{eq:HHcoms}. One obtains
\begin{equation}\label{Eq.HHbarphaseshift1}
\bar{\delta}_{00}=\bar{\delta}_{01\pm}=\bar{\delta}_{0J*}=\frac{\pi}{2},\quad\bar{\delta}_{10}=\bar{\delta}_{11\pm}=\bar{\delta}_{1J*}=0,
\end{equation}
or
\begin{equation}\label{Eq.HHbarphaseshift2}
\bar{\delta}_{00}=\bar{\delta}_{01\pm}=\bar{\delta}_{0J*}=\bar{\delta}_{10}=\bar{\delta}_{11\pm}=\bar{\delta}_{1J*}=\frac{\pi}{2},
\end{equation}
where $J=0,1,2$.
Then the solutions of the LECs read:
\begin{equation}
C_{0a}=-\frac{2\pi}{\mu\bar{\Lambda}},\quad C_{0b}=C_{1a}=C_{1b}=0,
\end{equation}
or
\begin{equation}
C_{0a}=C_{1a}=-\frac{2\pi}{\mu\bar{\Lambda}},\quad C_{0b}=C_{1b}=0,
\end{equation}
where $\bar{\Lambda}$ denotes the cut-off for the $D^{(*)}\bar{D}^{(*)}$ system.

\begin{table}[tb]
\centering
\caption{Partners of the $X(3872)$ predicted by HQSS or the two solutions of entanglement suppression given in Eqs.~\eqref{Eq.HHbarphaseshift1} and \eqref{Eq.HHbarphaseshift2}. The symbol ``$\odot$'' denotes the input $X(3872)$, ``$\otimes$'' represents its predicted partners, ``$\oslash$'' indicates no near-threshold state is allowed, and ``$-$'' signifies that no prediction can be made without further inputs. Moreover, ``$\oplus$'' means that the corresponding meson pair needs to be mixed with another one to get a spin partner of $X(3872)$, see Eqs.~\eqref{0++} and \eqref{1+-}.}
\begin{ruledtabular}
\begin{tabular}{p{6em}cccccc}
\multirow{2}{*}{Channel} & \multicolumn{2}{c}{HQSS} & \multicolumn{2}{c}{Eq.~\eqref{Eq.HHbarphaseshift1} predictions} & \multicolumn{2}{c}{Eq.~\eqref{Eq.HHbarphaseshift2}  predictions} \\
\cmidrule(lr){2-3}\cmidrule(lr){4-5}\cmidrule(lr){6-7}
& $\quad I=0 \quad$ & $\quad I=1 \quad$ & $\quad I=0 \quad$ & $\quad I=1 \quad$ & $\quad I=0 \quad$ & $\quad I=1 \quad$ \\
\midrule
$D\bar{D}(0^{++})$ & $\oplus$ & $-$ & $\otimes$ & $\oslash$ & $\otimes$ & $\otimes$ \\ 
$D\bar{D}^*(1^{++})$ & $\odot$ & $-$ & $\odot$ & $\oslash$ & $\odot$ & $\otimes$ \\ 
$D\bar{D}^*(1^{+-})$ & $\oplus$ & $-$ & $\otimes$ & $\oslash$ & $\otimes$ & $\otimes$ \\ 
$D^*\bar{D}^*(0^{++})$ & $\oplus$ & $-$ & $\otimes$ & $\oslash$ & $\otimes$ & $\otimes$ \\ 
$D^*\bar{D}^*(1^{+-})$ & $\oplus$ & $-$ & $\otimes$ & $\oslash$ & $\otimes$ & $\otimes$ \\ 
$D^*\bar{D}^*(2^{++})$ & $\otimes$ & $-$ & $\otimes$ & $\oslash$ & $\otimes$ & $\otimes$ 
\end{tabular}
\end{ruledtabular}
\label{HHbartab}
\end{table}

In both scenarios, we conclude that $X(3872)$ should have five spin partner states, all of them being isoscalar states, like the $X(3872)$ itself, so there are totally six weakly bound states in the isospin-$0$ channels of $D^{(*)}\bar{D}^{(*)}$ scattering. Also, it is noted that HQSS predicts only three isoscalar spin partners in the strict heavy-quark limit~\cite{Hidalgo-Duque:2013pva,Baru:2016iwj}, one $D^*\bar{D}^*$ state with $J^{PC}=2^{++}$~\eqref{Xp} and two mixing states with $J^{PC}=0^{++}$ and $J^{PC}=1^{+-}$:
\begin{align}
\label{0++}
J^{PC}=0^{++}:&\quad\cos
\frac{\pi}{6} |D\bar{D},I=0\rangle +\sin \frac{\pi}{6} |D^*\bar{D}^*,I=0\rangle,\\
\label{1+-}
J^{PC}=1^{+-}:&\quad\cos \frac{\pi}{4}\left|\frac{1}{\sqrt{2}}\left( D\bar{D}^*+D^*\bar{D} \right),I=0\right\rangle +\sin \frac{\pi}{4} \left|D^*\bar{D}^*,I=0\right\rangle.
\end{align} 
Therefore, as in the double-charm case, entanglement suppression again enlarges the symmetry for the spin degree of freedom of the light quarks from SU(2)$\times$SU(2) to SU(4), see Eqs.~\eqref{Eq.HHbarphaseshift1} and \eqref{Eq.HHbarphaseshift2}, predicting more states than HQSS.

Furthermore, if Nature chooses the solution in Eq.~\eqref{Eq.HHbarphaseshift2}, there would be six isovector hadronic molecules in addition, as listed in Table~\ref{HHbartab}. 
Two of these isovector states have quantum numbers $J^{PC}=1^{+-}$, and thus are in line with the existence of $Z_c(3900)$~\cite{BESIII:2013ris, Belle:2013yex} and $Z_c(4020)$~\cite{BESIII:2013qmu, BESIII:2013mhi} near the $D\bar{D}^*$ and $D^*\bar D^*$ thresholds, respectively. 
The existence of two $I(J^{PC})=1(1^{+-})$ molecular states likely implies the existence of in total six isovector states, as first proposed in Refs.~\cite{Bondar:2011ev, Voloshin:2011qa} based on HQSS in the analysis of the spin structure of the $Z_b(10610)$ and $Z_b(10650)$ discovered by the Belle Collaboration~\cite{Belle:2011aa}. 
Voloshin also proposed that the existence of both $Z_b(10610)$ and $Z_b(10650)$ could imply a light-quark spin symmetry~\cite{Voloshin:2016cgm}.
Here we have shown that by imposing entanglement suppression only a single $Z_c$ or $Z_b$ state with $J^{PC}=1^{+-}$ is needed as input to select the solution in Eq.~\eqref{Eq.HHbarphaseshift2}, which predicts five more isovector charmonium-like or bottomonium-like molecular states.

\section{Large-$N_c$ perspective}\label{LN}

In the real world, the gauge symmetry of QCD is $\text{SU}(3)$. One can treat the number of colors, $N_c$, as a parameter and study gauge theories for any $N_c$, and in particular the large-$N_c$ limit, where intriguing results can be obtained~\cite{tHooft:1973alw,Witten:1979kh}. 
Here we analyze the $1/N_c$ scaling of the  $D^{(*)}D^{(*)}$ and $D^{(*)}\bar{D}^{(*)}$ scattering amplitudes. 

In the large-$N_c$ limit, a quark propagator is depicted by a straight line, with each quark carrying a color index, symbolizing the flow of color. On the contrary, gluons in the adjoint representation of $\text{SU}(N_c)$ carry two color indices. As $N_c$ approaches infinity, the difference between $N_c^2-1$ and $N_c^2$ can be neglected so one may represent gluons like a quark-antiquark pair as double lines. Moreover, for closed color lines, the indices are unrestricted by initial and final states, allowing them to be of any color. Summing over all possibilities, a closed color line contributes a factor of $N_c$. Moving on to the behavior of the QCD coupling constant $g$ in the large-$N_c$ limit, it is crucial to maintain color confinement without alteration while keeping $\Lambda_\text{QCD}=\mathcal{O}(N_c^0)$, which necessitates choosing the coupling constant $g_s$ to be of $\mathcal{O}(N_c^{-{1}/{2}})$~\cite{tHooft:1973alw,Witten:1979kh}. 
According to the Lehmann-Symanzik-Zimmermann reduction formula, the $1/N_c$ scaling of a scattering amplitude may be obtained by counting the power of $N_c$ for a correlation function and dividing it by those of the amplitudes creating a meson from vacuum by an interpolating operator, such as $\langle D^{(*)+}|c\bar{d}|\Omega\rangle$. Such an amplitude scales as $\mathcal{O}(N_c^{1/2})$.

\begin{figure}[tb]
    \centering
    \includegraphics[width=0.75\textwidth]{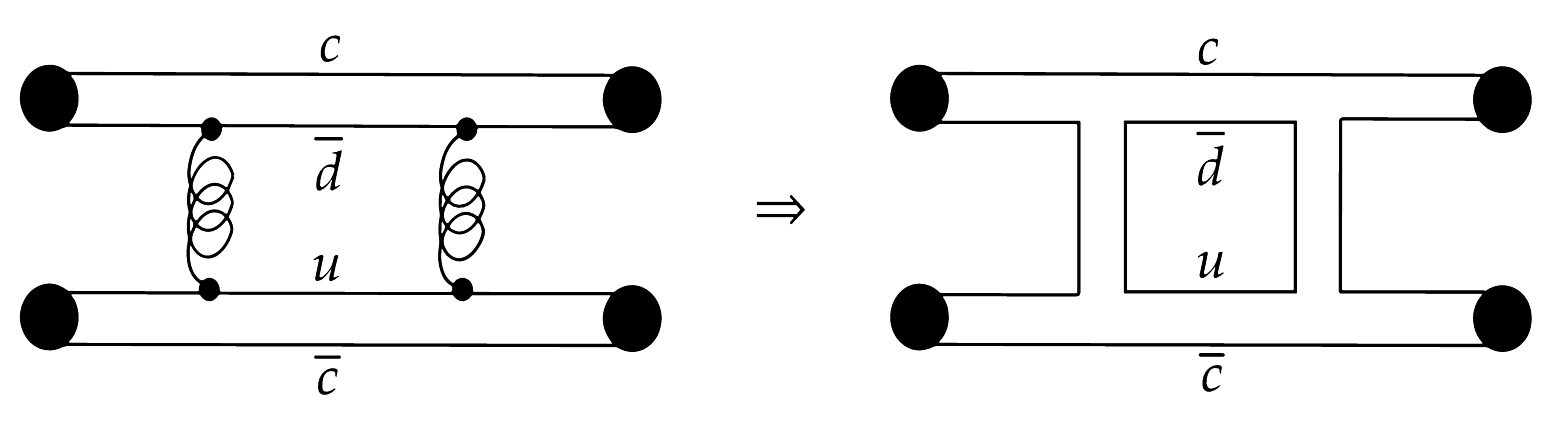}
    \caption{One of the LO diagrams for $D^{(*)+}\bar{D}^{(*)0}$ scattering (left), and the corresponding diagram with the double-line representation of gluons (right).}
    \label{fig: HHbar large-N limit}
\end{figure}

For the $D^{(*)}\bar{D}^{(*)}$ sector, let us analyze the scattering amplitude in the isovector $D^{(*)+}\bar{D}^{(*)0}$ channel. In the heavy-quark limit, we neglect the contribution from exchanging charm quarks. One diagram at LO of $1/N_c$ counting is shown in Fig.~\ref{fig: HHbar large-N limit}. 
In the large-$N_c$ limit, there are two closed color loops, four gluonic vertices, each contributing a factor of $g_s$, and four $\langle D^{(*)+}|c\bar{d}|\Omega\rangle$ or $\langle \bar{D}^{(*)0}|\bar{c}u|\Omega\rangle$ factors to be taken out. Hence, the order of the amplitude is
\begin{equation}
T^{I=1}(D^{(*)}\bar{D}^{(*)})=\mathcal{O}\left( N_c^2\times \left( N_c^{-{1}/{2}} \right) ^4\div \left( N_c^{{1}/{2}} \right) ^4 \right)=\mathcal{O}(N_c^{-2}).
\end{equation}
This amplitude is $1/N_c$ more suppressed than the isoscalar $D^{(*)}\bar{D}^{(*)}$ amplitude, in line with the Okubo-Zweig-Iizuka rule~\cite{Okubo:1963fa, Zweig:1964ruk, Iizuka:1966fk}. The isoscalar scattering can proceed through annihilating and creating light quark-antiquark pairs, and thus the LO color line diagrams without gluonic vertices consist of only one closed color loop. Correspondingly, one has
\begin{equation}
	T^{I=0}(D^{(*)}\bar{D}^{(*)})=\mathcal{O}\left( N_c \div \left( N_c^{{1}/{2}} \right) ^4 \right)=\mathcal{O}(N_c^{-1}). \label{eq:Nc_HHbar_I0}
\end{equation}

\begin{figure}[tb]
    \centering
    \includegraphics[width=0.75\textwidth]{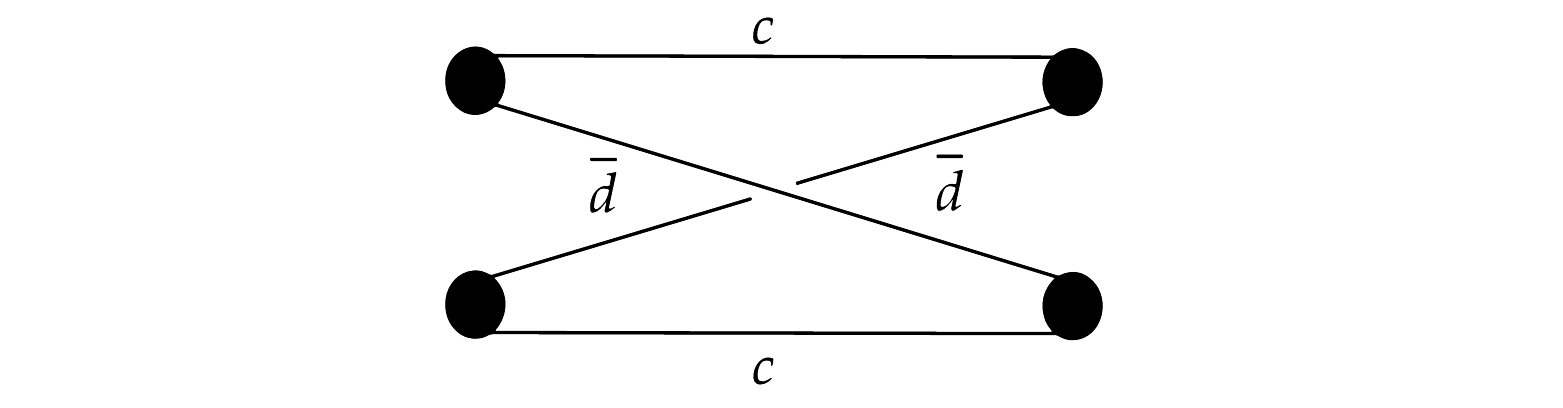}
    \caption{One LO diagram for $D^{(*)+}{D}^{(*)+}$ scattering.}
    \label{fig: HH large-N limit}
\end{figure}

The situation in the $D^{(*)}D^{(*)}$ sector is different, since the two charm mesons can always interact through exchange of light antiquarks no matter whether the total isospin is 0 or 1, see Fig.~\ref{fig: HH large-N limit}.
Therefore, the $1/N_c$ scaling of the scattering amplitude is similar to that in Eq.~\eqref{eq:Nc_HHbar_I0},
\begin{equation}
T^{I=0,1}(D^{(*)}D^{(*)})=\mathcal{O}\left( N_c\div \left( N_c^{{1}/{2}} \right) ^4 \right)=\mathcal{O}(N_c^{-1}).
\end{equation}

One sees that the $1/N_c$, together with HQSS and the $X(3872)$ input, would lead to the same scenario as in Eq.~\eqref{Eq.HHbarphaseshift1}. 
However, the existence of $Z_c(3900)$, $Z_c(4020)$ and $Z_b(10610)$, $Z_b(10650)$ implies subleading $\mathcal{O}(N_c^{-2})$ contributions in the $1/N_c$ expansion might be important for the scattering of a heavy-antiheavy meson pair.\footnote{In Ref.~\cite{Dong:2021lkh}, it has been shown that the interaction between two $J/\psi$ through the exchange of soft gluons, the amplitude of which also scales as $\mathcal{O}(N_c^{-2})$, might be strong enough to form a molecular state.} As such, the solution in Eq.~\eqref{Eq.HHbarphaseshift2} is beyond consequences of the large-$N_c$ limit.
Moreover, HQSS constrains that the LO $D^{(*)}D^{(*)}$ or $D^{(*)}\bar{D}^{(*)}$ interaction strengths for each isospin depends on two LECs (for instance, $C_{0a}$, $C_{0b}$ for $I=0$ and $C_{1a}$, $C_{1b}$ for $I=1$ in Eq.~\eqref{eq:LECs_C}). 
Without introducing more detailed dynamics, the large-$N_c$ limit does not provide a connection between these two LECs.\footnote{In Refs.~\cite{Dong:2021juy, Dong:2021bvy}, using a light-vector meson exchange model, the authors find that the $D^{(*)} \bar{D}^{(*)}$ potentials are the same for a given isospin; the pattern also holds for the $D^{(*)}D^{(*)}$ potentials. When the $\rho$ and $\omega$ meson masses are set to be equal, which holds in the large-$N_c$ limit~\cite{Witten:1979kh}, the isovector $D^{(*)}\bar D^{(*)}$ potentials obtained there vanish, consistent with the large-$N_c$ results. }
On the contrary, entanglement suppression predicts that the interaction strengths for each isospin are the same.

\section{Summary and discussion}\label{SUM}

In this study, inspired by the findings of previous works~\cite{Beane:2018oxh,Beane:2021zvo,Low:2021ufv,Liu:2022grf}, we studied consequences of entanglement suppression in the low-energy  $D^{(*)}D^{(*)}$ and $D^{(*)}\bar{D}^{(*)}$ scatterings.
These processes are currently of high interest due to the  discoveries of the $X(3872)$ and $T_{cc}(3875)^+$, which are proposed to be hadronic molecules of $D\bar{D}^*$ and $D^*D$, respectively. 
Using the $X(3872)$ and $T_{cc}(3875)^+$ as inputs, we found that entanglement suppression results in the same interaction strengths for all $D^{(*)}D^{(*)}$ pairs with the same isospin, unless the interaction is forbidden due to Bose-Einstein statistics; the same also holds for $D^{(*)}\bar{D}^{(*)}$. The isoscalar channels are at the unitary limit, thanks to the inputs of $X(3872)$ and $T_{cc}(3875)^+$, and molecular states are predicted as listed in Tables~\ref{HHtab} and \ref{HHbartab}. 
The isovector channels would be uncertain, being either noninteracting or at the unitary limit, corresponding to the two possible fixed points of two-body nonrelativistic scattering by a short-range potential. 
However, for the $D^{(*)}\bar{D}^{(*)}$ pairs, the existence of a single $Z_c(3900)$ or $Z_c(4020)$ (or $Z_b(10610)$ or $Z_b(10650)$ in the bottomonium sector) state with $J^{PC}=1^{+-}$ allows one to select the solution with all isovector channels at the unitary limit. In this case, entanglement suppression together with HQSS predicts five more isovector states around the $D^{(*)}\bar D^{(*)}$ thresholds, see Table~\ref{HHbartab}.

The spin symmetry of the light degrees of freedom from HQSS is $\text{SU}(2)\times\text{SU}(2)$, and it is now enlarged to \text{SU}(4) as the interaction between the heavy mesons does not depend on the total angular momentum of the light degrees of freedom, which is referred to as the light-quark spin symmetry, first proposed in Ref.~\cite{Voloshin:2016cgm}.\footnote{Such a scenario is realized in the light-vector-meson exchange model considering only vector couplings of the light vector mesons to heavy hadrons~\cite{Dong:2021juy,Dong:2021bvy}.} Therefore, we conclude that entanglement suppression does result in an emergent symmetry, which is light-quark spin SU(4) symmetry for systems of a pair of $S$-wave heavy mesons, in addition to the inherent HQSS in the low-energy sector of heavy mesons.

The predictions made here need to be contrasted with future experimental data or lattice QCD results, in order to test the validity of the conjecture connecting entanglement suppression to the origin of emergent symmetries made in Ref.~\cite{Beane:2018oxh}.

\begin{acknowledgments}

	We would like to thank Xiang-Kun Dong for useful discussions. This work is supported in part by the Chinese Academy of Sciences under Grants No.~YSBR-101 and No.~XDB34030000;
	by the National Key R\&D Program of China under Grant No. 2023YFA1606703;
	by the National Natural Science Foundation of China (NSFC) under Grants No. 12125507, No. 12361141819, and No. 12047503; and by NSFC and the Deutsche Forschungsgemeinschaft (DFG) through the funds provided to the Sino-German Collaborative Research Center CRC110 ``Symmetries and the Emergence of Structure in QCD'' (DFG Project-ID 196253076).
 
\end{acknowledgments}

\appendix

\section{Entanglement measure}\label{app}

This appendix presents a brief discussion of entanglement measure. In particular, we will show that the linear entropy employed in this paper is semi-positive definite and vanishes only on tensor-product states.

A general density matrix is defined as
\begin{equation}
\rho=\left|\psi\right>\left<\psi\right|,\quad \left|\psi\right>=\sum_{i,j}a_{ij}\left|i_{1}\right>\otimes\left|j_{2}\right>,
\end{equation}
and the normalization of $\left|\psi\right>$, $\left<\psi|\psi\right>=1$, means
\begin{equation}
\sum_{i,j}a_{ij}^{}a_{ij}^*=\text{Tr}\left(AA^\dagger \right)=1.
\end{equation}
where $(A)_{ij}=a_{ij}$. Since $H=AA^\dagger$ is a semi-positive definite Hermitian matrix, $\text{Tr}\left(H\right)$ is simply the sum of all its eigenvalues which are all real and $\geq0$, $\sum_{i}\lambda_i$.

The reduced density matrix can be obtained by tracing over subsystem 2:
\begin{equation}
\rho_1=\operatorname{Tr}_2(\rho)=\sum_{i,j,k}a_{ik}^{}a_{jk}^*\left|i_1\right>\left<j_1\right|.
\end{equation}
Then one can calculate
\begin{equation}
\begin{aligned}
\text{Tr}_1[\rho_1^2]&=\sum_{i,j,k,l}a_{ij}^{}a_{kj}^{*}a_{kl}^{}a_{il}^{*}\\
&=\text{Tr}\left(AA^\dagger AA^\dagger\right)=\sum_{i}\lambda_i^2,
\end{aligned}
\end{equation}
and therefore prove the semi-positivity of the linear entropy:
\begin{equation}
E(\left| \psi \right>) = 1 - \text{Tr}_1[\rho_1^2]=\left(\sum_{i}\lambda_i\right)^2-\sum_{i}\lambda_i^2\geq0,
\end{equation}
where we have used the fact that the eigenvalues of a density matrix sum up to unity.
The equality holds only when one of the $\lambda_i$ is equal to 1 and the rest are 0, which means that $A$ can be decomposed into the product of two vectors, $A=UV^T$, or equivalently, $a_{ij}=u_iv_j$, and this suggests $\left| \psi \right>$ should be a tensor product:
\begin{equation}
\left|\psi\right>=\left(\sum_{i}u_{i}\left|i_{1}\right>\right)\otimes\left(\sum_{j}v_{j}\left|j_{2}\right>\right).
\end{equation}

\bibliography{refs}

\end{document}